\documentclass[a4paper, 12pt]{article} 
\usepackage{english} 
\usepackage{epsfig}
\begin{document}
\title{Development of a Scintillating Fibre Detector for HERA-B}
\author{H. B. Dreis, F. Eisele$^*$, M. Hildebrandt, \\
  B. Schmidt, A. Sponsel, M. Ziegler 
\vspace{.5cm}  \\ \it
Physikalisches Institut der Universit\"at Heidelberg \\ Philosophenweg 12,
  D 69120 Heidelberg \rm }

\maketitle
\begin{abstract}
A fibre detector with multianode PM readout for high rate applications was developed for the HERA-B spectometer to cover the area near the beampipe. Such a
detector has to operate at particle rates up to $2 \star 10^4 \ mm^{-2}s^{-1}$
and has to tolerate a radiation dosis up to 1 Mrad/y. The light output of test
detectors with different scintillating   fibres  was measured as well as the light transport through clear fibres to the PM  as a function of radiation dosis and different conditions of irradiation. The geometry of the detector is described and results for a full size prototype are presented. 
\end{abstract} 
\section{Introduction}
The HERA-B experiment, presently under construction at DESY, Hamburg, is
designed to measure CP violation in the B-system. Neutral B-Mesons are produced by interactions of 820  GeV protons on a stationary (carbon) target. Only about
1 out of $10^6$ interaction produces a $B^0-\bar B^0$ pair. HERA-B
therefore plans a total interaction rate of 40 MHz leading to very high particle fluxes and high radiation levels in the detector. The particle flux is about $10^6 \ cm^{-2}
s^{-1}$ at 6 cm vertical distance $r_p$ from the beam and drops roughly as $1/r_p^2$. The innermost part of the tracking system ( inner tracker ITR) which covers a range $|x|<20$ cm and $|y|<25$ cm is therefore a big challenge for
 the tracking detector technology. The detectors have to
sustain a particle flux up to $2*10^6 \ cm^{-2}s^{-1} $ and a radiation dosis up to 1 Mrad/year.\par
The  baseline design for the inner tracker of HERA-B used Microstrip Gas Chambers (MSGC), where 4 L-shaped detectors covered one detector plane around the 
beam pipe. This detector technology faced very severe 
operation problems in intense hadron beams as first shown in a beam test at 
PSI \cite{PSItest}. Heavily ionising particles induce gas discharges which
destroy the electrodes if operated at useful gas gains of about 2000. Since
this problem could not be solved fast, we were forced to also investigate 
alternative detector concepts. The use of scintillating fibre detectors
for the inner tracker looked as an interesting possibility and was therefore followed up to a full size prototype.
\section{Geometry of a fibre tracking detector for HERA-B }
The following requirements are necessary to satisfy the needs of HERA-B .
 
\begin{figure}[ht] \begin{center}
\epsfig{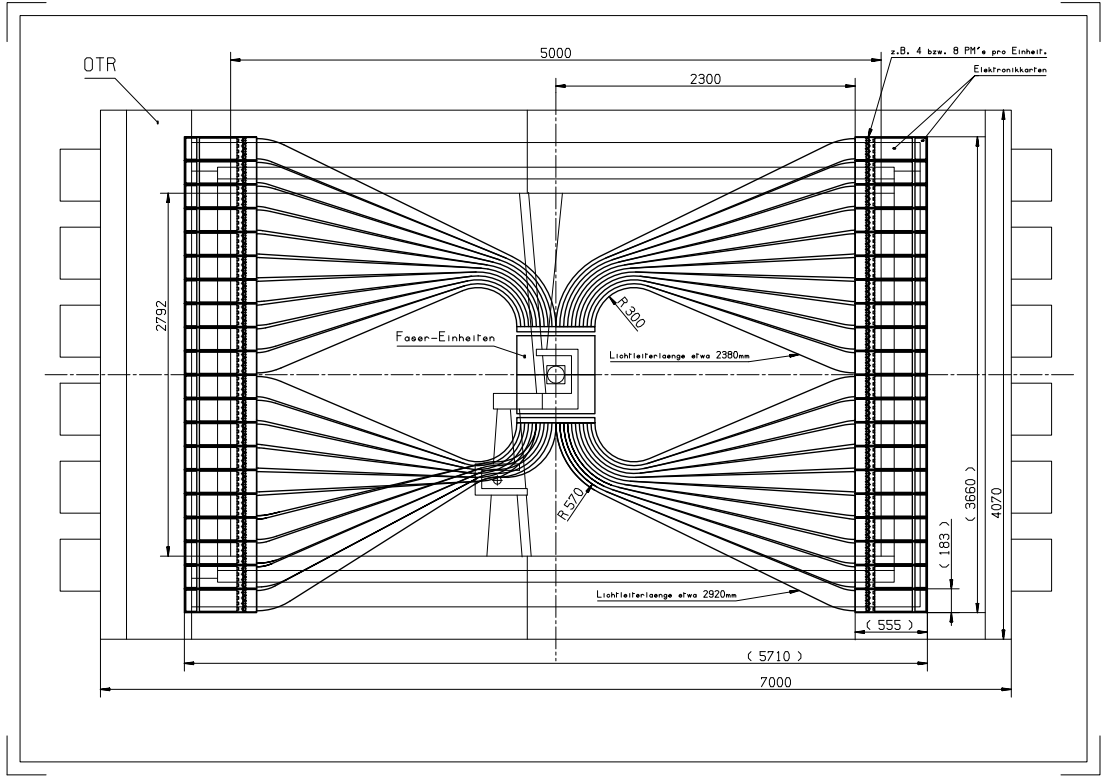} \end{center}
\parbox[t]{14cm}{\caption{\label{MSGCgeom}Typical  geometry of a fibre detector station for HERA-B . The fibre detector (small rectangle at the center) covers an area of about $44 \ cm \star 56 \ cm$ around the beampipe. The light is transported via clear fibres of up to 3 m length to the photomultipliers which are positioned to the left and right outside of the tracking region.}} 
\end{figure}

\begin{figure}[ht] \begin{center}
\epsfig{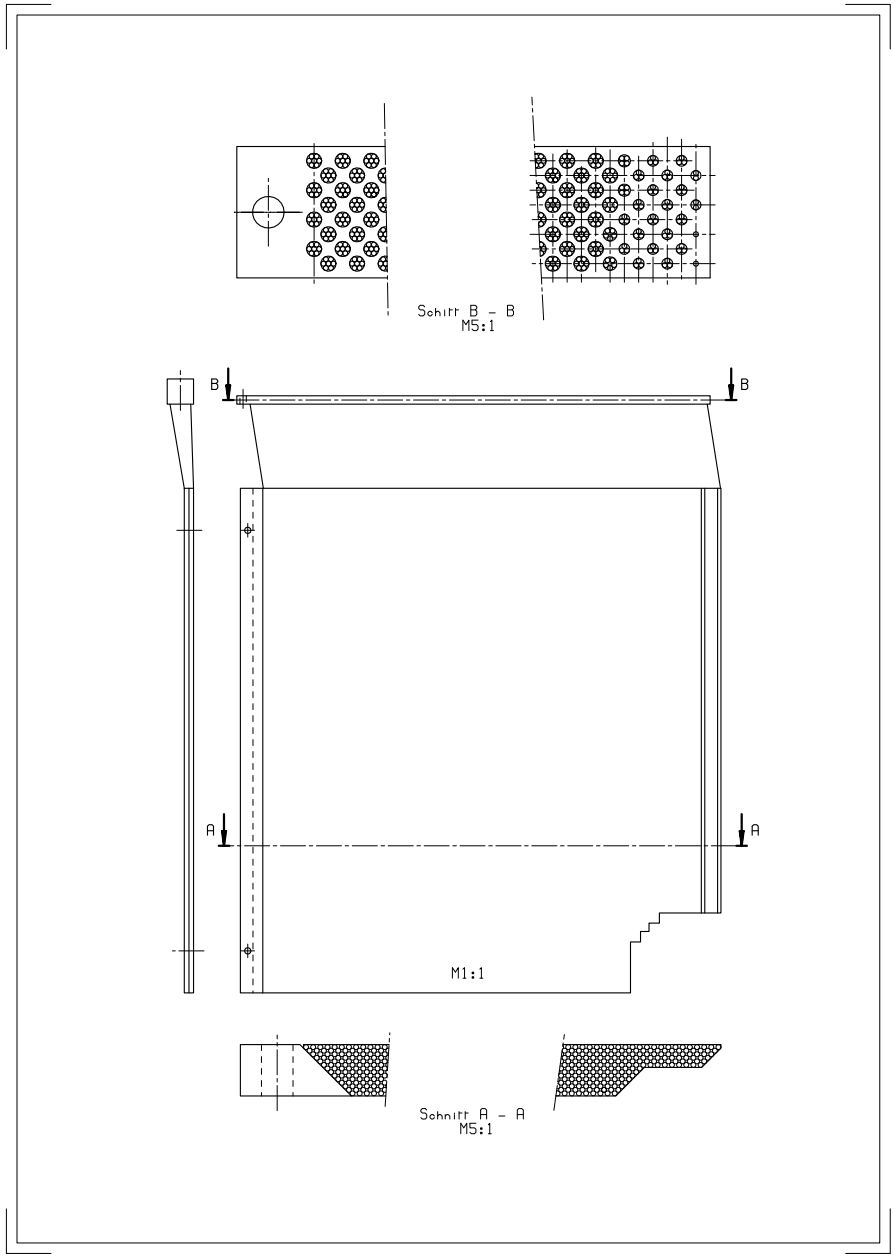} \end{center}
\parbox[t]{14cm}{\caption{\label{fibergeom}Geometry of a fibre detector for HERA-B . The insert on top shows the connector with the couplings of 7 scintillating fibres to one clear fibre. The insert at the bottom shows a cross section through the detector with the fibre geometry.}}
\end{figure}

\begin{itemize}
\item A detector plane needs $\ge $ 95\% efficiency for a minimum ionising particle.
\item The occupancy of one detector element should not exceed 6\% at 40
MHz interaction rate. This requires a granularity of readout of $\le 350 \ 
\mu m$.
\item The active area should be identical to the original MSGC geometry in
order to fit to the overall detector concept.

\item We need fast readout. The analog readout needs a pulse shaping of $\le \ 
100 \  ns$. In addition the detector has to deliver level one trigger signals for
each bunch crossing (every 100 ns).  We therefore opted for multianode
photomultipliers.
\item The scintillating light of each detector element has to be transported to
the outside of the sensitive tracking area up to the photomultipliers over distances between 1.0 to 3.5 m.

\end{itemize}
Figure \ref{MSGCgeom} shows a typical geometry for one fibre tracking station
with the arrangement of PM's and clear fibre light guides.
Figure \ref{fibergeom} shows the fibre geometry of the detector which was chosen
to meet these requirements. The detector is buildt from double clad fibres
with an outer diameter of .48 mm. In order to guarantee high efficiency, 
7 fibres which are positioned behind each other along the beam direction form
one readout road. They are coupled to a single clear fibre (1.7 mm diameter)
and are read out by one photomultiplier pixel. Neighbouring fibre roads overlap
by choosing a pitch of 340 $\mu m$. This guarantees that a minimum ionising 
particle with  angle $\le 100 \  mr$ w.r.t. vertical impact traverses at least
2.5 mm of scintillator. One detector plane has then 704 readout roads.
The scintillating fibres are coupled to clear fibres using  custom designed
connectors made of acrylic plates.\par
Compact multianode photomultipliers (Hamamatsu R5900-M64) were forseen for photon detection. These 14 stage tubes house 64 anodes with pixel sizes 
at the cathode of $2*2 \  mm^2$
and have a bialkali photocathode. This tube was specifically developed for the HERA-B  project. 
The clear fibre diameter of 1.7 mm is the maximum possible for this PM to
match it to one pixel and  to keep optical cross talk of neighbouring pixels below 1
\%. The 7 fibres within one road are therefore also the maximal number which can be read out by one pixel.
\section{Estimate of light yield}
\begin{figure}[ht] \begin{center}
\epsfig{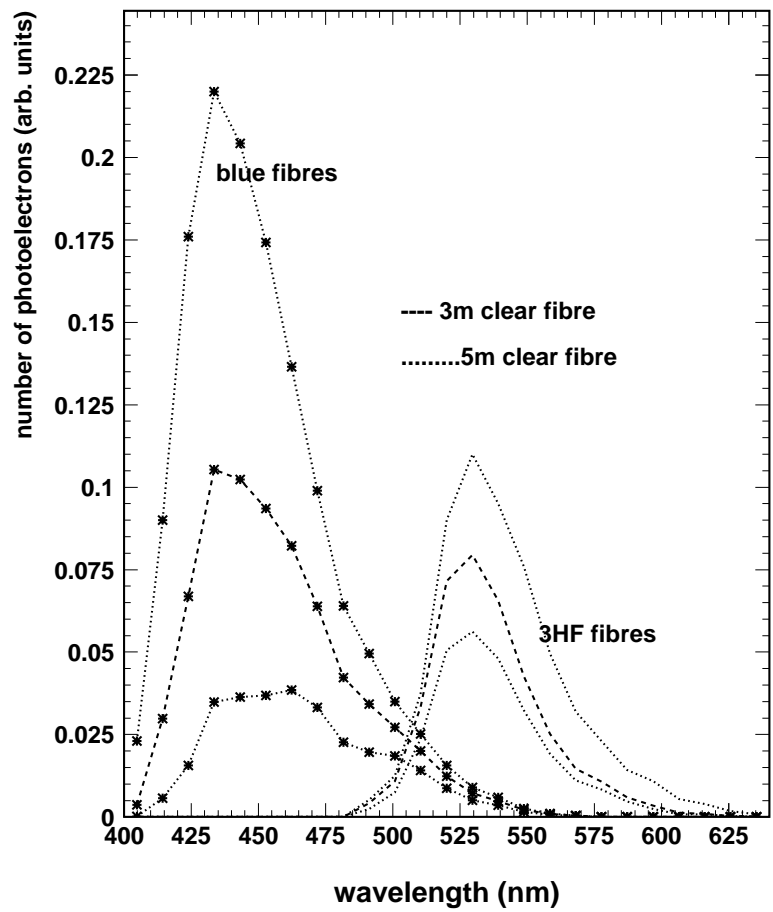} \end{center}
\parbox[t]{14cm}{\caption{\label{lightyield} Estimated number of
    photoelectrons vs. wavelength  for a blue and green fibre
    detector  for 3 different lengths of clear fibres (0.1 m,3 m and 5 m).}}
\end{figure}

In order to get an efficiency above 95\% for a minimum ionising particle we have to detect at least 3.0 photoelectrons in one road in average assuming we can rely  on 
single photoelectron events or 4.7 photoelectrons if we need at
least 2 photoelectrons for safe
particle detection. This light level has to be guaranteed over the life time of
the experiment, e.g. up to a total radiation dosis of about 3 Mrad for the
innermost part. \\
The following options were considered and experimentally tested.
\begin{itemize}
\item  \bf scintillating fibres:\rm \ We tested double clad scintillating fibres from three companies 
(Kuraray, Bicron and PolHiTech) and different emission wavelength. Blue fibres with maximum emission around 450 nm are best matched to the quantum efficiency 
of the bialkali photocathode but blue light is rather strongly absorbed in the clear fibres
especially after exposure to radiation.
Green (3HF) fibres with maximum emission around 540 nm are known to be more
radiation hard \cite{D0}, green light shows small absorption in the clear fibres but is poorly matched to the cathode efficiency. Finally we have also tested  fibres in the blue-green
emission region $(\sim 490 $ nm).
\item \bf clear fibres: \rm Double clad fibres of all three companies were tested.
Transmission of the light from  different scintillators was tested as a function of fibre length and radiation dosis.
\item The scintillating fibres have a typical length of 35 cm. In order to increase the light yield and minimize absorption effects we tested different reflectors at the fibre end.
\item The optical couplings between scintillating and clear fibres and to the PM were systematically tested.

\end{itemize}
Figure \ref{lightyield} shows the relative  number of photoelectrons for new
fibres  which was calculated on the basis of
published data for fibres and photomultiplier after 0.1 m, 3 m and 5 m of clear
fibre. Blue and green fibres were expected to deliver more than 15 resp.
9
photoelectrons  after 3 m of clear fibre which gave hope that such a detector could deliver enough light
even after radiation damage. Green fibres looked promising because it was
expected that the light yield though smaller at the beginning would be more
robust against irradiation. A green fibre  solution is very  attractive if one uses a
photon detector with high quantum  efficiency at long wavelength as used e.g. by D0
\cite {D0}.
Unfortunately it turned out to be impossible to produce multianode
photomultipliers with green enhanced  photocathodes with sufficient homogenity.

\section{ Measurements of relative light yield as a function of radiation dosis}
The relative light yield for the HERA-B  geometry and readout was measured in
the laboratory  using electrons from a
strontium source for  different combinations of scintillating and clear
fibres before and after irradiation .
\subsection{Experimental setup}
 Fibre bundles of the correct geometry but only  3 
scintillating fibres in one readout  row and with a width of typically 
2 cm were assembled using different glueing techniques. The length of the
scintillating fibres was 35 cm, they were glued together over a length of
about 25 cm, the typical length of a HERA-B  detector. The three  fibres in
a row were then coupled to a clear fibre (typically 3 m long) and read out
by  a multianode photomultiplier (Hamamatsu R5900-16) which had the same
quantum efficiency as the PM forseen for the final readout. The Sr source was
collimated and high energy electrons were selected by a twofold coincidence of
narrow scintillator counters below the fibre bundle. This arrangement
guaranteed that the electrons pass through the fibre bundle but not
necessarily within one road. We therefore coupled  the clear fibres of 
7 neighbouring roads
to  a single pixel. The pulseheight was recorded  with a flash ADC
system for every single trigger. Since the number of photoelectrons was
typically small the analysis was done by determining the average number of
photoelectrons from the measured spectra by folding single electron spectra
using Poisson probability.   \par
The coupling between scintillating and clear fibres was done with specially
designed connectors which allowed reproducible  light coupling to
better than 7\%. It was therefore possible to measure the relative light yield
of different fibre bundles
or  of one  fibre bundle after different steps of irradiation reliably using
the same setup of fibres  and readout channels. \par
The irradiation of fibres was done using a strong $ Co^{60}$  $\gamma $ source of BASF company,
Ludwigshafen. This source allows a radiation dosis of 50 krad/h at a distance of 5 cm and  is surrounded by a large irradiation
volume. By placing the fibres at  a proper distance from the source the dosis per unit time
could be adjusted from 1.2 Mrad per day to 10 krad per day. This possibility was e.g. used to
position  the clear fibres in such a way that they were irradiated according
to the expected  relative dosis along
their length in HERA-B . The light yield was not recorded during irradiation but 
could normally only be measured about 2 hours after the end of irradiation.
\subsection{ Production of fibre bundles}
The fibre bundles were produced by hand using a machined Al template with
grooves at a distance of 740 $\mu m$ which defined the position of the first
layer of scintillating fibres. Successive layers were then
positioned by the layer below. The fibres were glued together using standard
acrylic water based glue except for the fibre ends which were mostly embedded
in epoxy glue for better machining. This simple procedure produced rather
 regular
structures. Different types of glueing techniques have
been used.   For our standard fibre bundles the fibres were glued together
with transparent acrylic glue only in few  narrow strips perpendicular to the
fibres. This guarantees that air can access the fibres during irradiation.
We also  produced bundles where the fibres were
completely embedded in white acrylic glue. This white
glue prohibits any optical cross talk between the fibres within the bundle and
also changes to some extent the overall light yield.  
\subsection{Tests of mirrors and connectors}
In a first step we tested the reproducibility and effectiveness of connectors
and mirrors. The connection between scintillating and clear fibres is done 
using an air gap. The use of optical grease was also tested but improved the
light yield by only 10\% which does not justify the other risks. Both types of
fibres were glued into the inner  tube of a brass connector using epoxy glue.
The surface at the fibre ends was then milled and polished. The two
parts of the connector could then be screwed together guaranteeing a well
defined position and distance of the fibres. The transmission of these
connectors was determined to be greater than 80\% and this transmission was
reproducible to better than 5\%. \par
At the end of the fibre bundle (opposite to the readout) the scintillating
fibres were glued together using epoxy glue and then milled.
The absorption within the 35 cm of fibre length is not negligible especially
after irradiation. We therefore planned to use  a mirror to the end. The best
mirror turned out to be a sheet of aluminium which was coupled to the end of
the bundle using optical grease. Polishing of the fibre ends did not lead to
improvements. Using this simple method we reached good reflectivity resulting
 in an increase of light output by a factor 1.75
which turned out to be well reproducible. 
\subsection{ Relative light yield of scintillating fibres as a function of
  radiation dosis}
Various fibre bundles were produced using the same glueing technique and were
coupled to the PM using always the same  clear fibre bundle (Kuraray). 
Table \ref{fibres} gives a summary  about the materials which have been used.

\begin{table}\begin{center}
\caption{\label{fibres}Summary of double cladded fibres which have been used in the tests}
\vspace{0.5cm}
\begin{tabular}{l|l|l|l} \hline
\bf  company & \bf  fibre & \bf wavelength & \rm \\ \hline
& \bf scintillating fibres \rm & & \\ \hline
Kuraray & SCSF-78M & $\sim $ 450 nm & radiation hard \\
Bicron & BCF12 & $\sim $ 440 nm & \\
PolHiTech & 0246B & 450 nm & \\
\hline
Kuraray & 3HF(1500)M & 540 nm & radiation hard \\
Kuraray & PMP450 & 450 nm & test sample \\
Bicron & BCF 60 & 540 nm & \\ \hline
 & \bf clear fibres \rm & & \\ \hline
Kuraray & clear fibre & & \\
Bicron & BCF98 
& & \\
PolHiTech & PolifiOP-2-120 & & \\
\hline
\end{tabular}
\end{center}
\end{table}

\begin{figure}[ht] \begin{center}
\epsfig{file=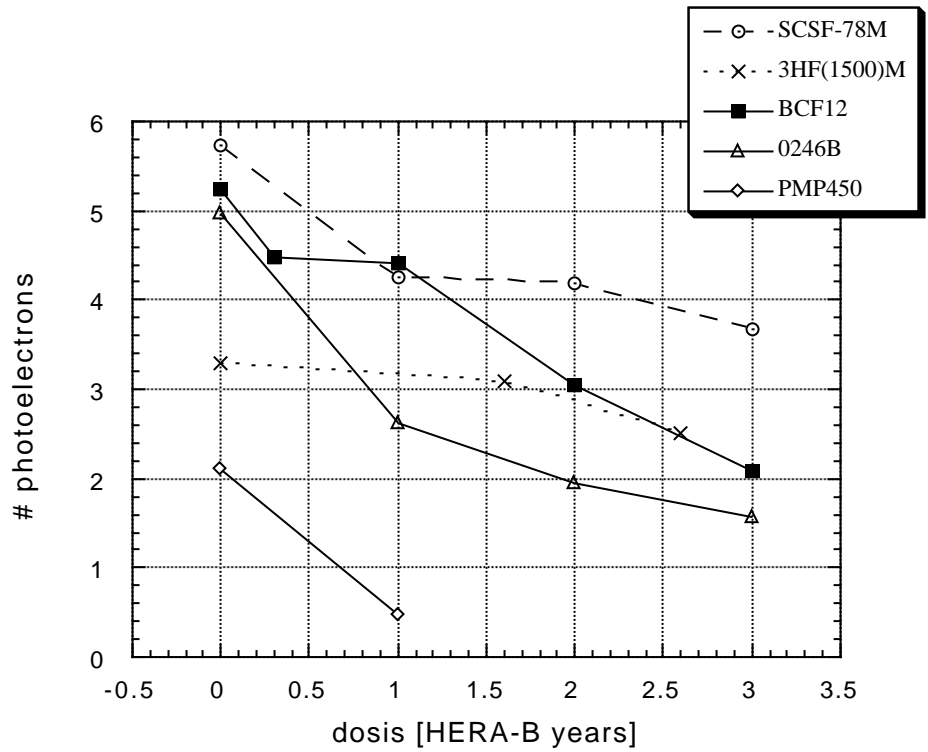,height=10cm,clip=} \end{center}
\parbox[t]{12cm}{\caption{\label{scintrad} Number of photoelectrons measured for 4 different scintillating fibre detectors versus radiation dosis. The typical error for each measurement is 7\%.
 One year of HERA-B  corresponds to about  1 Mrad. This dosis was  applied with a rate of about 50 krad/h.}}
\end{figure}

After several reference measurements the fibre bundles were  exposed to radiation starting with 300 krad
and then in steps of 1 Mrad  where this dosis was applied within 14 hours
e.g. at a  radiation level of about 50 krad/h (100 times faster than in HERA-B ). The light yield was measured after every step.
The results are shown in figure \ref{scintrad}.

 New fibres show very different 
number of photoelectrons. The best yield is obtained using Kuraray SCSF-78M
blue fibre followed by Bicron and PolHiTech blue fibres. Green 3HF fibres as
expected have a rather low number of photoelectrons  due to the low quantum efficiency 
of the PM for green light. The different fibres show rather different 
sensitivity to radiation. A common feature is  however, that a good fraction of the light loss appears already after 0.3 resp. 1 Mrad. Fibres with  PMP which were
produced as a test sample by Kuraray are destroyed completely after rather
small irradiation. Blue Kuraray fibres loose  about 36\% of their light yield after 3 Mrad, whereas the Bicron BCF12 fibres loose 66 \% and the PolHiTech fibres 0246B 68\%.
This light loss is entirely caused by a shortening of the absorption length in the
scintillating fibre as can be demonstrated by the change of light yield along
the length of the fibre bundle. The green 3HF fibres loose only 20\% and are
thus relatively robust against
radiation as expected and reported before \cite{D0}.\par
The light yield was monitored over a long period  to see if scintillating fibres recover from
radiation damage. Under our conditions of rather fast irradiation  no
recovery was observed over a period of typically 40 days  for fibre
bundles which were only partially glued and therefore allowed air circulation.
\subsection{Transparency of clear fibres }
\begin{figure}[ht] \begin{center}
\epsfig{file=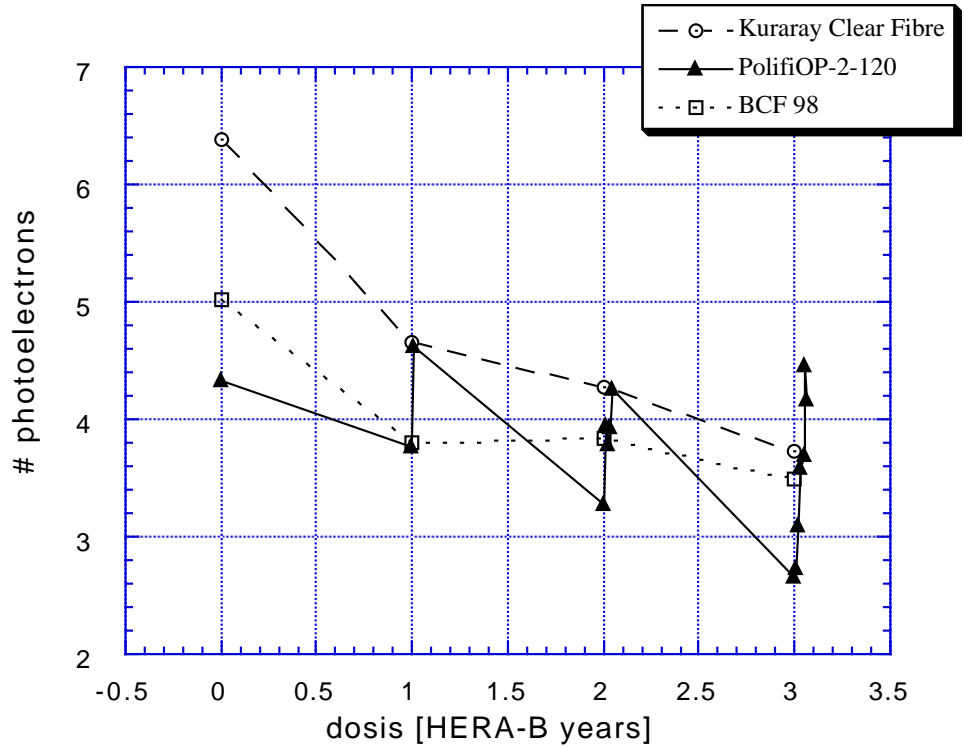,height=10cm,clip=} \end{center}
\parbox[t]{12cm}{\caption{\label{clearfibres}Number of photoelectrons measured
    for 3 different clear fibres of 3 m length  versus radiation dosis. After
    each irradiation step the light yield was measured several times to
look for recovery phenomena. Recovery of light
    yield was only  observed for the PolHiTech fibres over periods up to 70 hours.
 One year of HERA-B  corresponds to an irradiation which varies between 300 krad
 and 10 krad along the fibre. The scintillating light was produced by a blue
 Kuraray fibre detector with 3 fibres in one road.}}
\end{figure}

The relative transparency of clear fibres  from all three companies was measured
as a function of irradiation  using always the same  unexposed scintillating fibre bundle  
(Kuraray blue SCSF-78M fibres). The clear fibres were positioned in the irradiation volume in such a way that the 'natural' irradiation of HERA-B  was simulated
which starts with 300 krad/year at 30 cm distance from the beam pipe and comes
down to 10 krad/year at a radial distance of 3.0 m. The results are shown in figure \ref{clearfibres}. 
Clear Kuraray fibres have by far the best transparency for blue light before
irradiation. After an irraditation corresponding to 3 years of HERA-B  however they have lost 40\% of transparency and all three fibres are about equal on first sight. 
Kuraray and Bicron clear fibres show essentially no recovery from radiation damage. 
 In contrast the 
 PolHiTech fibres show very strong recovery 
with time scales up to 700 hours. These fibres can recover to full transparency if one waits long enough. Since they start however with a transparency which is only 2/3 of the Kuraray fibres they are finally more or less equally good. Again the loss of transparency is related to a change of absorption length in the clear fibres.\par
It was originally hoped that the use of green 3HF fibres would profit from
a better transparency of the clear fibres after irradiation. This is in principle true however not for our readout with photomultipliers. Green fibre bundles
in combination with clear fibres showed the same total loss in the number of photoelectrons as blue fibres. The reason for this is, that the light which is
transmitted up to the photocathode after irradiation is shifted to longer wavelengths which leads to a reduced quantum efficiency at the photocathode and therefore a reduced number of photoelectrons. This  compensates the advantage of better transparency in the clear fibres. Green fibres are therefore not a viable choice.

\section{ Radiation damage of fibres under varying conditions}
\begin{figure}[ht] \begin{center}
\epsfig{file=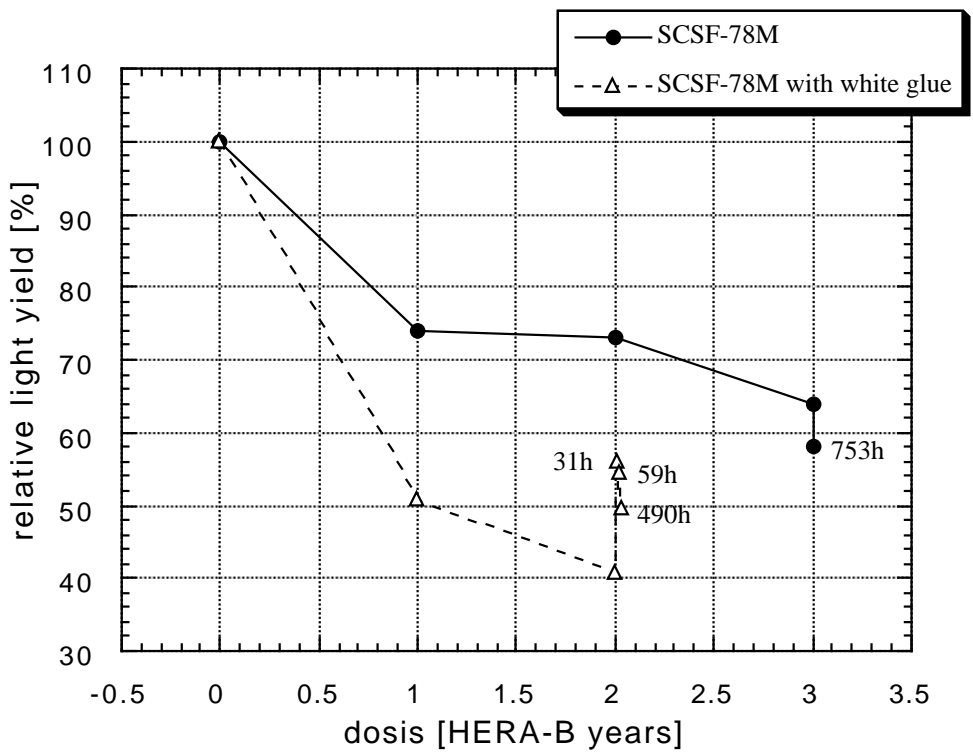,height=10cm,clip=} \end{center}
\parbox[t]{12cm}{\caption{\label{bavaria}Relative number of photoelectrons measured for two fibre detectors with blue Kuraray fibres versus radiation dosis. For the first detector the scintillating fibres were glued together only in narrow strips such that air could circulate through the fibres. For the second detector the scintillating fibres were fully imbedded in white acrylic glue. }}
\end{figure}

\begin{figure}[ht] \begin{center}
\epsfig{file=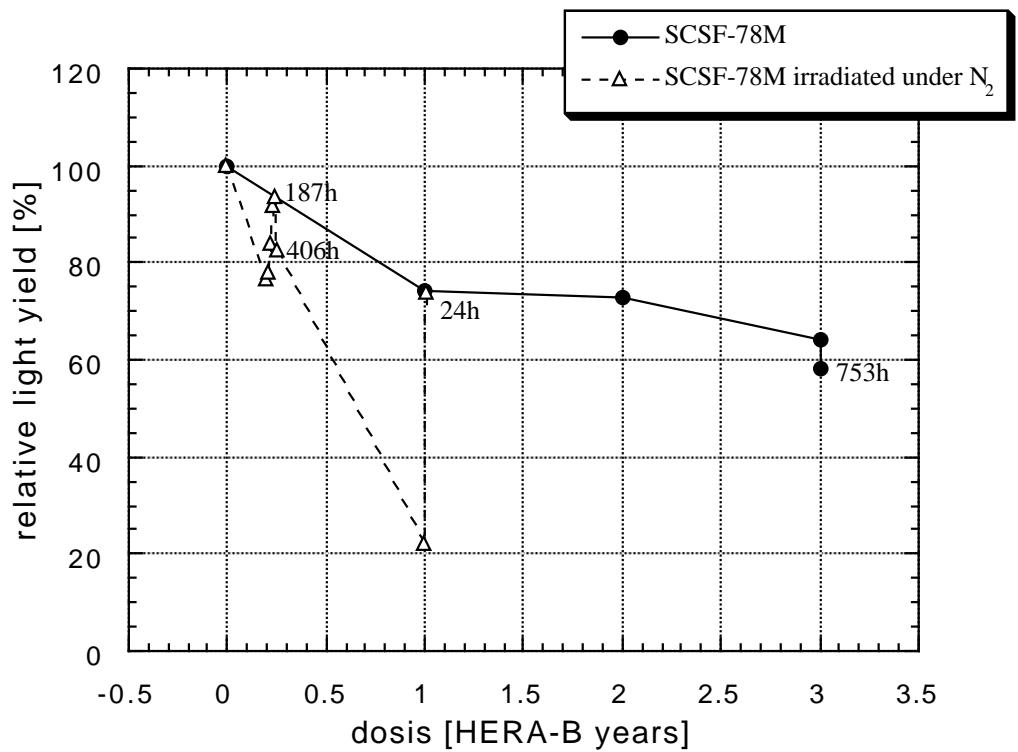,height=10cm,clip=} \end{center}
\parbox[t]{12cm}{\caption{\label{N2scint}Relative number of photoelectrons measured for two blue scintillating fibre detectors with partial glueing versus radiation dosis. One detector was exposed to air, the other to nitrogen during irradiation. After irradiation both detectors were exposed again to air  and their light yield was measured several times over a period of several days. }}
\end{figure}

The loss of light yield for a fibre detector at HERA-B  is expected to be very severe according to the results presented so far. We have therefore made auxiliary tests to
see if the radiation damage can be reduced or if better recovery can be obtained. These tests included different glueing techniques for the fibre detector,
irradiation in $N_2$ and studies of radiation damage as a function of dosis
rate.
Figure \ref{bavaria} compares the relative light yield of a standard fibre bundle using 
blue Kuraray fibres and partial glueing of the fibres with a bundle where the same fibres were fully imbedded in white arylic glue.
 The radiation damage for
the detector with white glue is substantially larger but it shows partial recovery from radiation damage over time periods of about  60 hours. Figure \ref{N2scint} compares the loss of light yield for two standard Kuraray SCSF-78M detectors (partially glued)
where one was irradiated under air the other in nitrogen atmosphere.
The light loss for the bundle which was exposed in $N_2$  is much larger if
measured immediately after irradiation. However if this bundle is exposed to air then the light yield partially recovers over time periods of about 24 hours to a value still lower than what  is obtained after irradiation in air.
These results clearly indicate that oxygen plays a major role for radiation damage and recovery. Under our conditions of relatively moderate dosis rates these 
results may be interpreted in the following way: for fibres which are exposed under air and where the fibres are easily accessed by air (fibres not fully embedded in glue) the radiation damage seems to recover partially already during irradiation.  These bundles therefore show no subsequent recovery. $N_2$ atmosphere 
does not allow recovery and therefore leads to stronger damage.
 Clear Kuraray fibres show no significant difference if irradiated in $N_2$
 atmosphere compared to air.
\par
We have finally made some tests to study how the radiation damage depends on the rate of irradiation. For Kuraray scintillating and clear fibres no rate dependence was observed for rates between 100 krad/day and 1 Mrad/day if the irradiation was in air. The PolHiTech clear fibres on the other hand demonstrated a very complicated behaviour. Even very modest irradiations led to a large loss of light yield which however recovered fully after some time. We have no safe way to judge how these fibres would behave under the slow irradiation in the HERA-B  experiment.
\section{ Prototype detector}
A full size prototype fibre detector was manufactured by Kuraray company using
SCSF-78M scintillating fibres with a diameter of .48 mm according to the
geometry of figure \ref{fibergeom}.
The fibres of one row (mostly 7 fibres) were bundled in a 1.5 mm hole of  the 
connector  which was produced from an acrylic plate. Optical inspections on a measuring table showed that the accuracy 
of
perpendicular fibre position over the whole area of the detector is better than $\sigma=1  \  \mu m$. This detector was also tested in an electron test beam \cite{Zeuthen} with rather good results. \par
    In conclusion the production of such a fibre detector is industrially
    possible with adequate quality.
\section{Beam test results and  absolute light yield for a HERA-B  fibre detector}
\begin{figure}[ht] \begin{center}
\epsfig{file=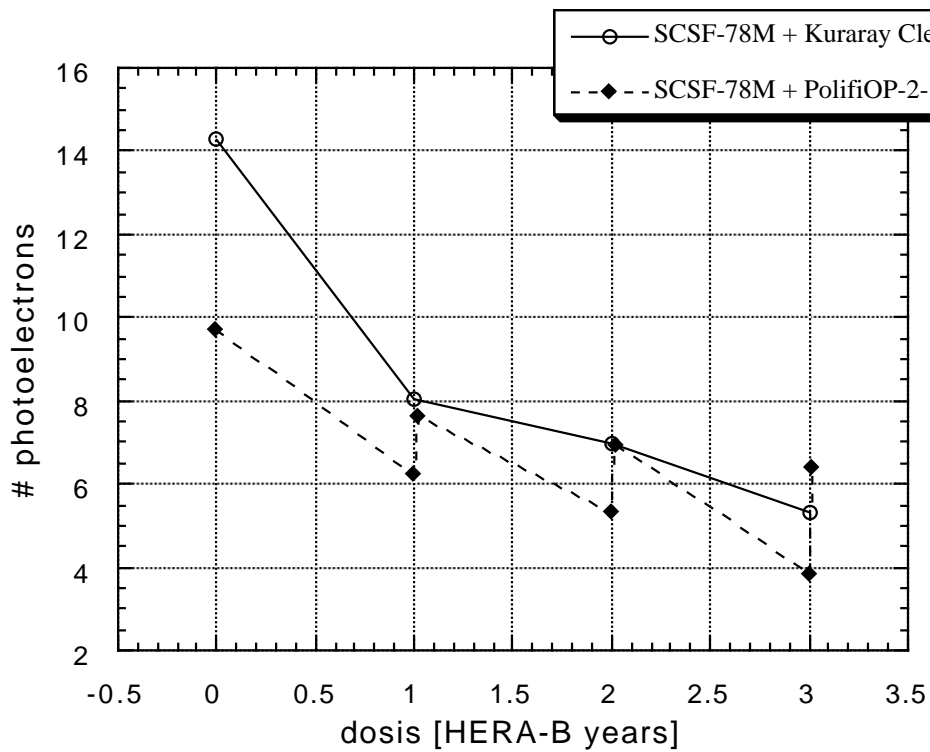,height=10cm,clip=} \end{center}
\parbox[t]{12cm}{\caption{\label{scint+clear} Expected number of photoelectrons for  a 
 fibre detector with blue Kuraray fibres with 7 fibres in one row  for two types of  3m long clear fibres versus radiation dosis. Both scintillating and clear fibres have been irradiated according to the expected radiation profile at HERA-B .The typical error of a measurement is 7\%.}}.
\end{figure}

The total  light yield for the best combinations of scintillating and clear
fibres  as measured in the laboratory are  shown in figure \ref{scint+clear} extrapolated to the full detector thickness of 7 fibres in a row. Whereas about 14 photoelectrons are expected for new fibres, this number is reduced to  5.5 after 3 years of operation. 
It should be noted that these numbers are upper limits for the number of
photoelectrons for the HERA-B  detector  because the
low energy electrons from the strontium source will normally scatter and thus have a longer effective path in the scintillating fibres compared to high energy minimum ionising particles.  It is therefore important to determine the absolute number of photoelectrons in a high energy test beam which also allows to
measure geometric effects like the sharing of light between neighbouring roads,
cross talk and the number of photoelectrons due to low energy background and Cherenkov light in the clear fibres.\par
These measurements were done in an electron test beam (5 GeV) at DESY 
by our collegues from Zeuthen \cite{Zeuthen} and 
in the HERA-B  experimental area. It turned out consistently that the laboratory measurements with a source strongly overestimated the number of photoelectrons.
A new fibre detector as forseen for HERA-B  using blue Kuraray SCSF-78M fibres and Kuraray clear fibres of 3 m length gave only 7 photoelectrons for one minimum ionising particle. After 3 years of operation this number could then drop to only
2.75 photoelectrons according to the radiation damage
 measurements. In addition it turned out that the number of fibres which
showed accidental signals was huge. These signals are essentially all single
photoelectron signals which are produced by soft particles, optical crosstalk
in the fibres and Cherenkov light in the clear fibres ($\sim $ 50\%). For
efficient tracking and the derivation of trigger signals we have therefore to
require at least 2 photoelectrons per minimum ionising particle. An efficiency
of the detector of $\ge $ 95\% under these conditions requires at least 4.0 photoelectrons per minimum ionising particle which cannot be guaranteed after irradiation.

\section{Conclusions} 
       Our measurements of light yield were done in a way to get  the most direct evaluation of a fibre detector as needed for HERA-B . More generic and systematic tests of the same materials have been carried out by our collegues at Zeuthen \cite{Zeuthen}.
They also answered the very important question if our results obtained with 
gamma ray irradiation are representative for the irradiation with hadrons and neutrons which would happen in the  HERA-B  experiment. For this purpose they 
compared radiation damage induced by proton and electron irradiation for the same dosis. It turned out that the long term damage is identical, $\gamma $
irradiation is therefore representative also for the radiation damage induced
by a hadron beam.\par
Our test results  show that a fibre detector as described in section
2 would satisfy the needs of the HERA-B  detector at the beginning. The expected
radiation damage of scintillating and clear fibres under  the conditions applied in  our tests is however so large that such a detector would have reduced efficiency already after one year of operation.
 It has to be pointed out however, that our results on radiation damage have
 been obtained by applying a rather large radiation dosis over a relatively short time period (typically 50 krad per hour). There is no proof that the same damage would occur at HERA-B  with a dosis rate of typically 
0.4 krad/hour. It cannot be excluded that the damage at HERA-B  would be substantially different.\par
Based on the available information we had to conclude however that 
safe operation over several years of HERA-B  running could not be
guaranteed. This fact in combination with the very high price tag for such a
detector led to the decision to abandon this technology for HERA-B .

\newpage
\section*{Acknowledgement}
We acknowledge the support of Dr. Seher and Mr. Mosbach from BASF company, Ludwigshafen, which allowed us to use the 
$Co^{60} \ $ source for  the
irradiations of our fibres. This work profited strongly from  collaboration with our
collegues from the Institut f\"ur Hochenergiephysik in Zeuthen \cite{Zeuthen}
in the area of fibre detectors and beam tests
and from the university Siegen  (A. Lange, U. Werthenbach, G. Zech and T. Zeuner) which carried out the tests of the multianode
photomultipliers. This project is  supported by the Ministerium f\"ur Bildung,
Wissenschaft, Forschung und Technologie, Bonn.

\newpage

\end{document}